\documentclass[3p]{elsarticle}
\usepackage{hyperref}
\usepackage{amssymb}
\usepackage{bm}
\usepackage{dcolumn}
\usepackage{amssymb}
\usepackage{amsmath}
\usepackage{graphicx}

\usepackage{epstopdf}
\usepackage{epsfig}

\usepackage{subfig}
\usepackage{booktabs}
\usepackage{multirow}
\usepackage{setspace}
\usepackage{array}
\usepackage{threeparttable}
\usepackage{txfonts}

\usepackage{exscale}
\usepackage{relsize}
\journal{Phys. Lett. B 820 (2021) 136546, \url{https://doi.org/10.1016/j.physletb.2021.136546}}

\begin{document}

\begin{frontmatter}



  \newcommand*{\PKU}{School of Physics and State Key Laboratory of Nuclear Physics and
    Technology, Peking University, Beijing 100871,
    China}
  \newcommand*{\CHEP}{Center for High Energy Physics, Peking University, Beijing 100871, China}
   \newcommand*{\CIC}{Collaborative Innovation Center of Quantum Matter, Beijing, China}

  \title{Pre-burst neutrinos of gamma-ray bursters accompanied by high-energy photons}

  \author[a]{Jie Zhu}
  \author[a,b,c]{Bo-Qiang Ma\corref{cor1}}

  \address[a]{\PKU}
  \address[b]{\CHEP}
  \address[c]{\CIC}

%  \address[d]{\CHPS}
  \cortext[cor1]{Corresponding author \ead{mabq@pku.edu.cn}}

  \begin{abstract}

Previous researches on high-energy neutrino events from gamma-ray bursters (GRBs) suggest a neutrino speed variation $v(E)=c(1\pm E/E^{\nu}_{\mathrm{LV}})$
with ${E}^{\nu}_{\rm LV}=(6.4\pm 1.5)\times10^{17}~{ \rm GeV}$, together with an intrinsic time difference ${\Delta {t}_{\rm in}=(-2.8\pm 0.7)\times10^2~{\rm s}}$, which
means that high-energy neutrinos come out about 300~s earlier than low-energy photons in the source reference system. Considering the possibility that pre-bursts of neutrinos may be accompanied by high-energy photons, in this work we search for high-energy photon events with earlier emission time from  100 to 1000~s before low-energy photons at source by analyzing Fermi Gamma-ray Space Telescope (FGST) data.
We perform the searching of photon events with energies larger than 100~MeV, and find 14 events from 48 GRBs with known redshifts.
Combining these events with a $1.07~\rm{TeV}$ photon event observed by the Major Atmospheric Gamma Imaging Cherenkov telescopes (MAGIC), we suggest a pre-burst stage with a long duration period of several minutes of high energy neutrino emissions accompanied by high energy photons at the GRB source.

  \end{abstract}

  \begin{keyword}
    pre-burst\sep gamma ray burst\sep light speed variation\sep neutrino speed variation\sep Lorentz invariance violation



  \end{keyword}

\end{frontmatter}

\section{Basic formulas for speed variation analysis}

Lorentz invariance is a basic assumption in Einstein's relativity. However, it is speculated from quantum gravity
that the Lorentz invariance might be broken at the Planck scale ($E_\mathrm{Pl}\simeq 1.22\times10^{19}~\mathrm{GeV}$).
Amelino-Camelia {\it et al.}~\cite{method1,method2} first suggest testing
Lorentz violation by comparing the arrival times between high energy and low energy particles from gamma-ray bursters (GRBs).
For a particle propagating in the quantum space-time with energy $E\ll E_{\rm Pl}$~, the Lorentz violation (LV) modified dispersion relation can be written in a general form as the leading term in Taylor series~
\begin{equation}\label{eq:1}
E^2\simeq p^2c^2+m^2c^4-s_nE^2\left(\frac{E}{E_{{\rm LV}, n}}\right)^n,
\end{equation}
where $E$, $p$, $m$ represent the energy, momentum and mass of the particle respectively, $c$ is the light speed in vacuum,
$s_n=\pm1$ indicates whether high-energy particles travel faster ($s_n=-1$) or slower ($s_n=+1$) than low-energy particles, $n=1$ or $n=2$ as usually assumed which shows the leading order of the influence from LV, and $E_{\rm{LV},n}$ represents the nth-order Lorentz violation scale. Since photons and ultra-high energy neutrinos are both ultra-relativistic
particles, it is reasonable to set $m=0$ in the discussion. Using the relation $v=\partial E/\partial p$, we can get
the modified velocity relation with energy $E$
\begin{equation}\label{eq:2}
v(E)=c\left[1-s_n\frac{n+1}{2}\left(\frac{E}{E_{{\rm LV},n}}\right)^n\right].
\end{equation}
With the expansion of the universe taken into consideration~\cite{formula}, the time difference from LV between two particles from the same source with energy of $E_{\rm{h}}$ and $E_{\rm{l}}$ is
\begin{equation}\label{eq:time}
\Delta t_{\rm LV}=s_n\frac{1+n}{2H_0}\frac{E_{\rm h}^n-E_{\rm l}^n}{E_{{\rm LV},n}}\int^z_0\frac{(1+z^\prime)^n {\rm d} z^\prime}{\sqrt{\Omega_{\rm m}(1+z^\prime)^3+\Omega_\Lambda}},
\end{equation}
where $z$ is the redshift of the source of the two particles, $H_0$, $\Omega_{\rm{m}}$ and $\Omega_{\rm{\Lambda}}$
are universe constants. Here we adopt the present day Hubble constant~\cite{pgb} $H_0=67.3\pm 1.2~ \rm{km~ s}^{-1}\rm{Mpc}^{-1}$,
the pressureless matter density~\cite{pgb} $\Omega_{\rm{m}}=\mathrm{0.315^{+0.016}_{-0.017}}$ and the dark energy density~\cite{pgb}
$\Omega_{\Lambda}=\mathrm{0.685^{+0.017}_{-0.016}}$. In this work, we focus on the circumstances of $n=1$, so Eq.~(\ref{eq:time}) can be expressed as
\begin{equation}\label{eq:4}
\Delta t_{\rm LV}=s(1+z)\frac{K}{E_{\rm LV}},
\end{equation}
where $s=\pm1$ is the sign factor and
\begin{equation}\label{eq:LVfactor}
K=\frac{E_{\rm h}-E_{\rm l}}{H_0 }\frac{1}{1+z}\int^z_0\frac{(1+z^\prime) {\rm d} z^\prime}{\sqrt{\Omega_{\rm m}(1+z^\prime)^3+\Omega_\Lambda}},
\end{equation}
is the LV factor. In our discussion, $E_{\rm{h}}$ ($\sim1~\rm{GeV}$ or higher) is much more higher than $E_{\rm{l}}$ ($\sim100~\rm{keV}$), so it is reasonable to take $E_{\rm{l}}$ as 0.

The observed arrival time difference $\Delta t_{\rm obs}$ between two particles detected on the Earth is actually caused by two factors: the LV time correction $\Delta t_{\rm LV}$ in the propagation and the intrinsic time difference $\Delta t_{\rm in}$ at the source~\cite{Ellis,shaolijing,zhangshu}. $\Delta t_{\rm in}$ is only related to the intrinsic mechanism of the GRB source. Hence we have
 \begin{equation}
 \Delta t_{\rm obs}=t_{\rm h}-t_{\rm l}=\Delta t_{\rm LV}+(1+z)\Delta t_{\rm in}.
 \label{eq:6}
 \end{equation}
 where $t_{\rm h}$ and $t_{\rm l}$ represent the arrival times of high-energy and low-energy particles. Considering Eq.~(\ref{eq:LVfactor}), we rewrite Eq.~(\ref{eq:6}) as
\begin{equation}
\frac{\Delta t_{\rm obs}}{1+z}=\Delta  t_{\rm in}+s\frac{K}{E_{\rm LV}}.
\label{eq:linear}
\end{equation}
 According to Eq.~(\ref{eq:linear}), there would be a linear relation between $\Delta t_{\rm obs}/(1+z)$ and $K$ if the energy dependence speed variation does exist, and particles emitted at a same intrinsic time in the source reference frame would fall on a straight line in the $K$-$\Delta t_{\rm obs}/(1+z)$ plot. The slope of the line represents $s/E_{\rm{LV}}$ and the intercept represents the intrinsic time difference $\Delta  t_{\rm in}$ between high-energy particles and low-energy photons.

 \section{Physics picture from previous researches}\emph{}

As sources with high-energy particle emissions, GRBs are natural laboratories for high energy physics, and the particles from GRBs have the potential to reveal novel features of the universe.
Because the energies of the particles from GRBs can be much more higher than energies of the particles from laboratories on the Earth, we can use them to test the predictions from
theoretical physics, e.g., the Lorentz violation. There are many researches on the tests of LV with high-energy particles from GRBs, and they focus on photons and neutrinos.

An attempt to test LV from high-energy photons was in Ref.~\cite{xu1}. In this work, it suggests a light speed variation of $v(E)=c(1-E/E^{\gamma}_{\mathrm{LV}})$ with $E^{\gamma}_{\mathrm{LV}}=3.6\times10^{17}~\mathrm{GeV}$ and an intrinsic time difference
$\Delta t_{\rm{in}}=(-10.7\pm1.5)~\rm{s}$, which means high-energy photons come out earlier than low-energy ones at the GRB source and the
speed of high-energy photons is slower than low-energy ones. In this physics picture, it suggests a pre-burst stage that high-energy photons come out about 10 seconds earlier than low-energy photons at the GRB source, and because of light speed variation, high-energy photons travel slower than low-energy photons, and the light speed difference and the long cosmological distances lead to an expectation that one usually observes low-energy photons earlier than high-energy photons, as is indeed the case in the earlier observations of GRB data. This conclusion was soon supported by a remarkable high energy event of GRB160509A in Ref.~\cite{xu2}. More recently, a direct evidence for the pre-burst stage is shown in Ref.~\cite{Jie} with the observation of
several pre-burst events which are earlier detected  than the prompt low energy photons.

Unlike photons, neutrinos are able to escape from dense astrophysical environments.
%and overcome the pair production problem which limits the photon energy.
Therefore ultra-high energy cosmic neutrinos can be observed on the Earth and provide a powerful tool to explore LV physics~\cite{Jacob:2006gn,Amelino-Camelia:2015nqa,Amelino-Camelia:2016fuh,Amelino-Camelia:2016ohi}.
Analysis on PeV and TeV GRB neutrinos~\cite{Huang1} suggests a neutrino speed variation of
$v(E)=c(1-s E/E^{\nu}_{\mathrm{LV}})$ with $E^{\nu}_{\mathrm{LV}}=6.5\times10^{17}~\mathrm{GeV}$ and $s=\pm1$ and an intrinsic time difference $\Delta t_{\rm{in}}=(1.7\pm3.6)\times 10^3~\rm{s}$. This suggestion shows that the speed of neutrinos can be both slower and faster than the light speed constant $c$, and the LV scale is not the same as the LV scale of photons but of the same order of magnitude. In the case of PeV GRB neutrinos, the time delay caused by LV
can be really large, e.g., the largest time difference in Ref.~\cite{Huang1} is 75 days, and it is much more larger than $10^3~\rm{s}$
as the order of magnitude of $\Delta t_{\rm{in}}$ determined by the data of PeV GRB neutrinos, so considering the uncertainty of the energy and the constants of the universe we can not expect an accurate value of
$\Delta t_{\rm{in}}$. In a later research on near-TeV GRB neutrinos from IceCube~\cite{Li}, it is suggested that
$E^{\nu}_{\mathrm{LV}}=(6.4\pm1.5)\times 10^{17}~\rm{GeV}$ and $\Delta t_{\rm in} =(-2.8\pm0.7)\times10^2~{\rm s}$, which
is more accurate and shows that high-energy neutrinos come out about $300~\rm{s}$ earlier from the GRB source than low-energy photons in the source reference frame.

Researches on high-energy photons and neutrinos reveal the following physics picture of GRBs. In the reference frame of the source GRB, about $300~\rm s$ before the low-energy photon burst, a pre-burst stage of high-energy neutrinos firstly happens. The ultra-high energies of neutrinos lead to big time lags or leads because of velocity variation, so we can observe ultra-high-energy neutrino events even 2 or 3 months later or before observing the GRB lights. Then about $10~\rm s$ before the low-energy photon burst, a pre-bust stage of high-energy photons happens. Different from neutrino events, high energies of the photon events only lead to time lags, so most of these events can be observed
after the low-energy photon burst and only few of them can be observed before the low-energy photon burst.
At the pre-burst stage of neutrinos, high-energy neutrinos can interact with the matters of the GRB source at the edge,
so these neutrino events can be accompanied by some high-energy photon events in the source reference frame. So to find the high-energy photon events at the pre-burst stage of neutrinos will be an indirect support for the physics picture of GRB neutrinos. However considering the expansion of the universe and the light speed variation, we can not simply search for
photon events just 300~s before a GRB. According to Eq.~(\ref{eq:6}), time difference at the source will be magnified by
a factor $(1+z)$, so we need to search for high-energy photon events with a larger time range.

\section{Data acquisition and analysis}

We search for high energy photon events from the Fermi Gamma-ray Space Telescope (FGST) data. FGST consists of the Fermi Large
Area Telescope (LAT)~\cite{LAT} and the Gamma-Ray Burst Monitor (GBM)~\cite{GBM}. LAT aims to collect high-energy events while GBM aims to collect low-energy events. The GBM data can be downloaded from the Fermi website~\cite{gbm_data} while the LAT data need to be retrieved and downloaded from this website~\cite{lat_data}. Considering that redshift plays an important role in our analysis, we focus on GRBs which are not only detected by FGST but also have redshifts recorded. In this work we search 48 GRBs from GRB~080916C to GRB~210204A. The trigger time $t_{\rm{trig}}$ of GBM is usually assumed as the onset time of GRBs, so in the following discussion we take $t_{\rm{trig}}$ as zero point of time.
Detailed discussion on the criteria of arrival time of low-energy photon events $t_{\rm l}$ is shown in Ref.~\cite{Liu}, however in this work the absolute value of arrival time of high-energy photon events $t_{\rm h}$ is much more larger than $t_{\rm l}$, so we simply take $t_{\rm l}=t_{\rm{trig}}=0$. For data of high-energy photon events from LAT, we need to specify an exact search range. We select LAT data with reconstructed energies in the $100~\rm{MeV}$-$100~\rm{GeV}$ range as Ref.~\cite{catalog1} did, and the lower limit is chosen to reject events with poorly reconstructed directions and energies.
Since the time difference at the source will be magnified by the factor $(1 + z)$, we set a larger time range to ensure that photons emit from -1000~s to -100~s in the source reference frame are selected. For the searching angle radius and background analysis, we still choose the same strategy as Ref.~\cite{catalog1} did. We use a fixed-radius circular region of interest set at $12^{\circ}$ to ensure that 99\% events are selected, and then we estimate the probability of each $\gamma$-ray being associated with the GRB by using the Fermi ScienceTool~\cite{sciencetool} \emph{gtsrcprob}, and then we choose events that have probabilities $>0.8$ of being associated with the GRBs.
The data are listed in Table~\ref{tab:events}. We notice that the Major Atmospheric Gamma Imaging Cherenkov telescopes (MAGIC)~\cite{magictelescope} found an ultra-high-energy photon events with energy 1.07~TeV and arrival time 73.6~s~\cite{magicdata1,magicdata2}, and the same analysis on the MAGIC event below suggests that it may come from the pre-burst stage of neutrinos, so we also list it in the table.

\begin{threeparttable}[t]
  \begin{centering}
    \caption{The data of observed pre-burst events}
    \begin{tabular}{p{30mm}<{\centering}p{10mm}<{\centering}p{14mm}<{\centering}p{14mm}<{\centering}p{15mm}<{\centering}p{15mm}<{\centering}p{24mm}<{\centering}p{10mm}<{\centering}}
      \hline
      \hline
      GRB        & $z$   & $t_{\rm h}$~(s)  & $E_{\rm obs}$~(GeV) & $E_{\rm source}$~(GeV)     &$\frac{\Delta t_{\rm obs}}{1+z}(\rm s)$     & $K(\times 10^{17}~ \rm{s}\cdot\rm{GeV})$ &$p$ \\
      \hline
      \text{090510A} & 0.903 & -1543.07 & 0.317 & 0.603 & -810.86 & 0.764 & 0.82 \\
      \hline
     \text{090902B} & 1.822 & -513.126 & 1.17 & 3.30 & -181.83 & 3.77 & 0.84 \\
     \hline
     \text{091003A} & 0.8969 & -1156.97 & 0.327 & 0.620 & -609.93 & 0.785 & 0.87 \\
     \hline
     \text{141028A} & 2.33 & -1230.13 & 0.109 & 0.362 & -369.41 & 0.372 & 0.80 \\
 %\text{141028A} & 2.33 & -2.7703 & 0.121 & 0.405 & -0.831922 & 0.416 & 0.87 \\
 %\text{141028A} & 2.33 & -0.013148 & 0.149 & 0.496 & -0.00394835 & 0.510 & 0.88 \\
 %\hline
 %\text{150403A} & 2.06 & -164.362 & 1.84 & 5.64 & -53.7131 & 6.14 & 0.90 \\
     \hline
     \text{170405A} & 3.51 & -1888.34 & 0.247 & 1.11 & -418.70 & 0.886 & 0.85 \\
     \hline
                    &      & -1004.15 & 0.332 & 0.794 & -420.15 & 0.979 & 0.85 \\
     \text{181010A} & 1.39 & -830.676 & 0.272 & 0.649 & -347.56 & 0.801 & 0.88 \\
                   &      & -780.3 & 1.28 & 3.06 & -326.49 & 3.78 & 0.93 \\
                    &      & -697.326 & 0.333 & 0.795 & -291.77 & 0.981 & 0.88 \\
     \hline
     \text{191011A} & 1.722 & -1048.13 & 0.614 & 1.67 & -385.06 & 1.95 & 0.88 \\
     \hline
                    &     & -398.586 & 0.301 & 0.624 & -192.55 & 0.792 & 0.88 \\
     \text{201021C} & 1.07 & -320.235 & 0.321 & 0.665 & -154.70 & 0.845 & 0.86 \\
                    &     & -294.255 & 0.129 & 0.267 & -142.15 & 0.339 & 0.83 \\
 %\text{201021C} & 1.07 & -141.71 & 0.294 & 0.608 & -68.4589 & 0.772 & 0.89 \\
% \text{201021C} & 1.07 & -87.2267 & 0.370 & 0.765 & -42.1385 & 0.972 & 0.94 \\
% \text{201021C} & 1.07 & -12.8385 & 0.469 & 0.972 & -6.20217 & 1.23 & 0.84 \\
     \hline
     \text{201221D} & 1.046 & -952.789 & 0.446 & 0.912 & -465.68 & 1.16 & 0.83 \\
      \hline
      \hline
      190114C by MAGIC &0.4245   &73.6   &$1.07\times 10^{3}$   &$1.52\times 10^{3}$   &51.667   &$1.577\times10^{3}$\\
      \hline
      \hline

    \end{tabular}%

    \begin{tablenotes}
      \item Data of high-energy photon events from FGST and MAGIC. $z$ is the redshift of the GRB, $t_{\rm h}$ is the arrival time of the event when $t_{\rm trig}$ is set to zero, $E_{\rm obs}$ is the energy of the the event observed by the telescope while $E_{\rm source}$ is its energy in the source GRB, and $p$ is the probability that the event is associated with the GRB calculated by the Fermi ScienceTool~\emph{gtsrcprob}. $\Delta t_{\rm obs}/(1+z)$ and $K$ are parts of Eq.~(\ref{eq:linear}), and because of $t_{\rm l}=t_{\rm trig}=0$, here $\Delta t_{\rm obs}=t_{\rm h}-t_{\rm l}=t_{\rm h}$. Data of $t_{\rm h}$ and $E_{\rm obs}$ is from LAT, and the references for the redshifts of the GRBs are \cite{090510Az}(090510A),\cite{090902Bz}(090902B),\cite{091003Az}(091003A),\cite{141028Az}(141028A),\cite{170405Az}(170405A),\cite{181010Az}(181010A),\cite{191011Az}(191011A),
          \cite{201021Cz}(201021C),\cite{201221Dz}(201221D),\cite{190114Cz1,190114Cz2}(190114C),
          and the data of the MAGIC event are from Ref.~\cite{magicdata2}.
    \end{tablenotes}
    \label{tab:events}%
  \end{centering}
\end{threeparttable}%
\\

\begin{figure}[!h]
  \centering
  \subfloat[]{
    \label{fig:1a}
    \begin{minipage}{1\textwidth}
      \centering
      \includegraphics[width=\textwidth]{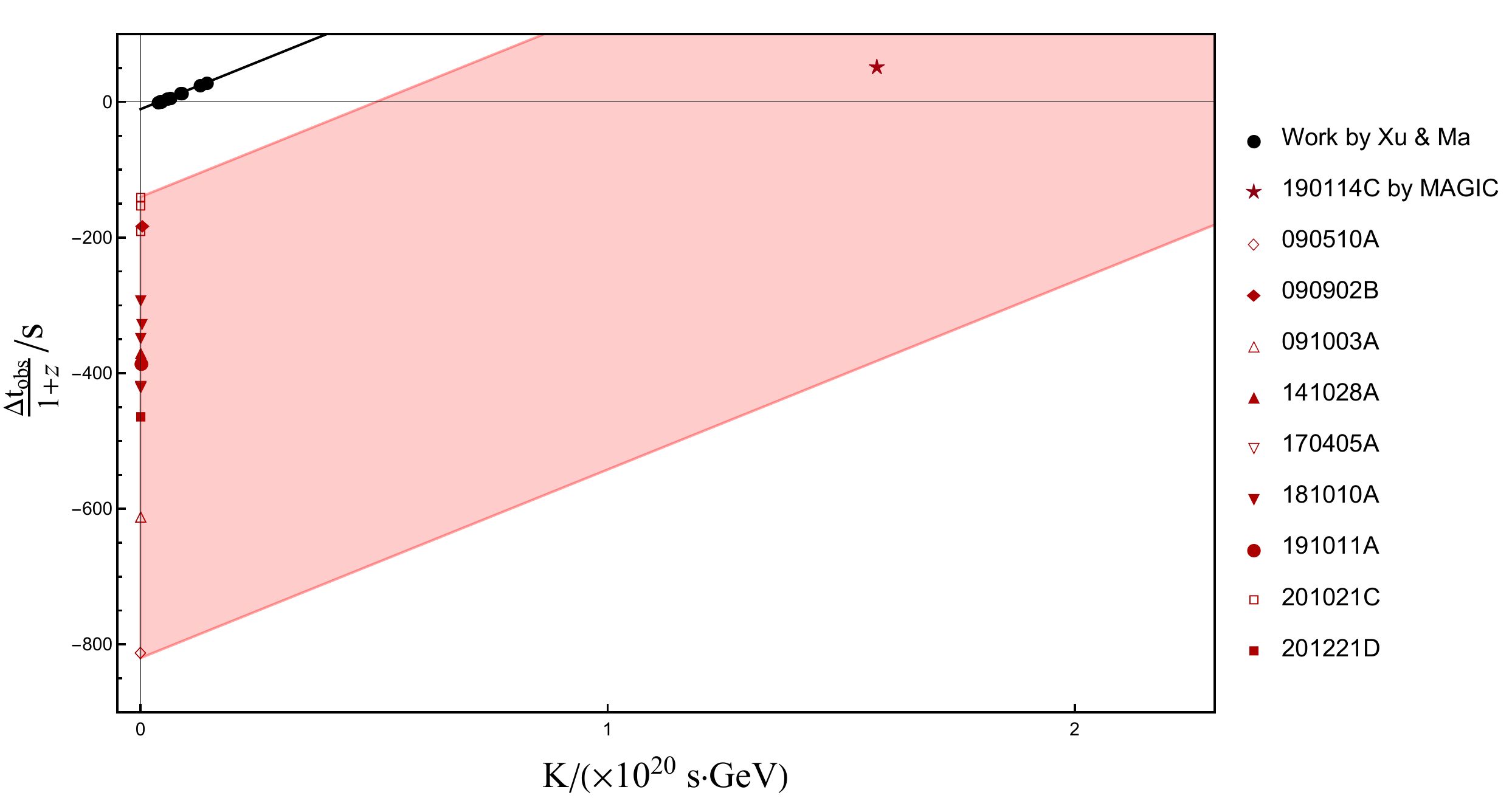}
    \end{minipage}
  }

  \subfloat[]{
    \label{fig:1b}
    \begin{minipage}{1\textwidth}
      \centering
      \includegraphics[width=\textwidth]{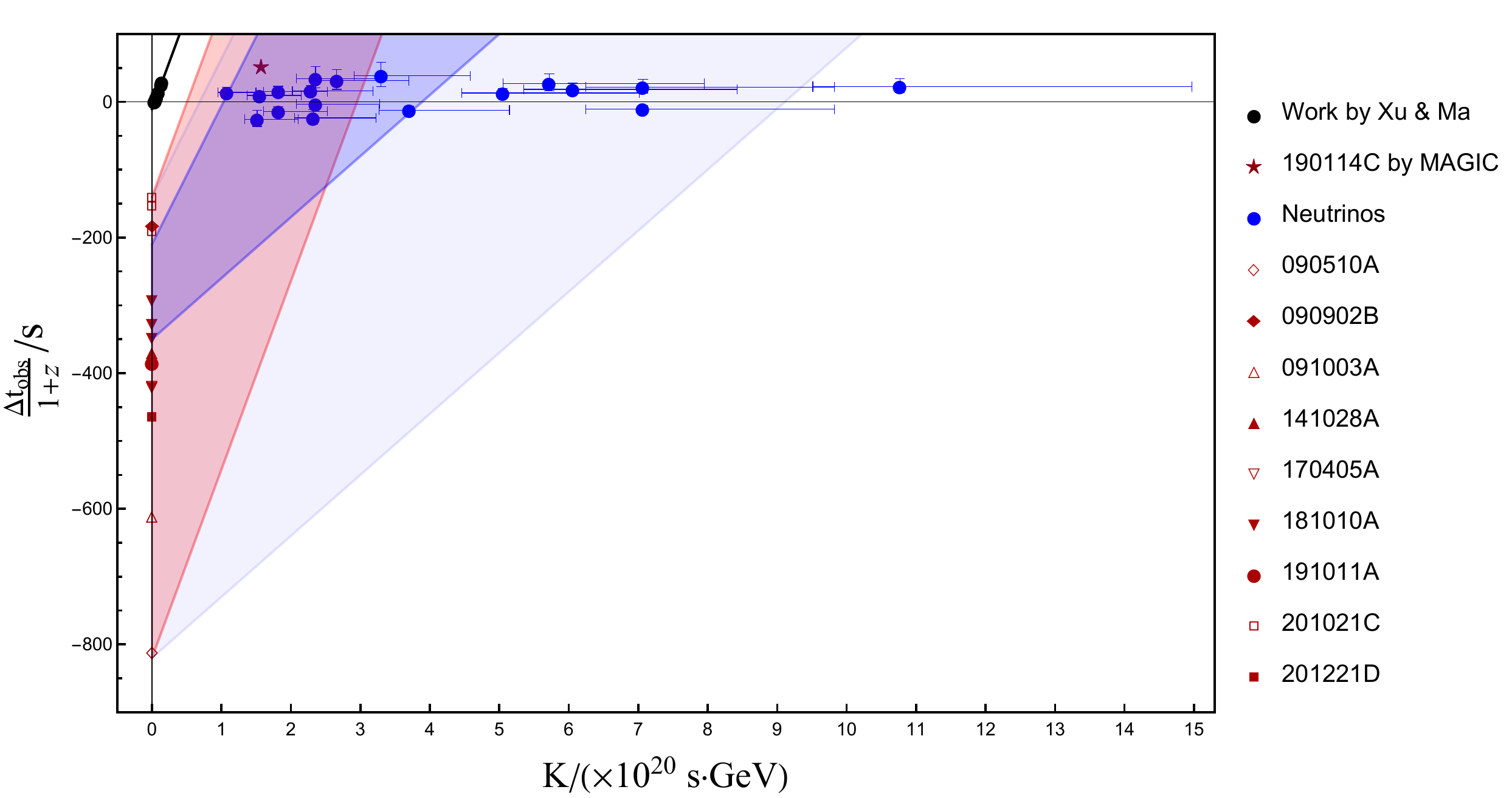}
    \end{minipage}
  }
  \caption{The $\Delta t_{\rm{obs}}/(1+z)$ versus $K$ plot for photon data in Table~\ref{tab:events} and in Refs.~\cite{xu1,xu2} (denoted as Work by Xu~\&~Ma), and for neutrino data in Ref.~\cite{Li}. In the upper panel~(a) we draw events in Table~\ref{tab:events} and in Refs.~\cite{xu1,xu2} and in the lower panel~(b) we add neutrino events. In both figures the black line is the mainline of Ref.~\cite{xu1} as reference, and the edges of the red~(dark) area share the same slope of the black line, and the intercepts of the two edges are -140~s and -820~s. In the lower panel the deep blue area is the maximum envelope of the fitting result in Ref.~\cite{Li}, and the light blue area keeps the same slope
  of the neutrino events~\cite{Li} but intercepts changed to -140~s and -820~s.   \label{fig:1}}
\end{figure}

Combining with data of high-energy photon and neutrino events from Refs.~\cite{xu1,xu2,Li}, we draw these events in the $K$-$\Delta t_{\rm obs}/(1+z)$ plot shown in Fig.~\ref{fig:1}. In the upper panel Fig.~\ref{fig:1a} we only include photon events from Refs.~\cite{xu1,xu2}, and in the lower panel Fig.~\ref{fig:1b} we also add near-TeV neutrino events from Ref.~\cite{Li}. Drawing a straight line through a point with slop $1/E^{\gamma}_{{\rm{LV}}}$ from Refs.~\cite{xu1,xu2}, we can get $\Delta t_{\rm in}$ from the intercept which represents the time difference between the event and low-energy photon bursts in the source reference frame. In Fig.~\ref{fig:1a}, we draw a red area with the same slope from Refs.~\cite{xu1,xu2} and intercepts from -140~s to -820~s, we can see that all of the events fall inside the area, including the 1.07~TeV photon event observed by MAGIC. This figure suggests that the events from FGST with negative arrival time are related with the 1.07~TeV event from MAGIC with positive arrival time, and they all come from the same time frame before the low-energy photon bursts of GRBs. In Fig.~\ref{fig:1b}, we add near-TeV neutrino events from Ref.~\cite{Li} as blue points. The redshifts of these neutrino GRB sources are unknown, and in Refs.~\cite{Huang1,Li} an approximate estimation of redshifts with errors is adopted, so here we draw these events with error bars. The deep blue area is the maximum envelope of the fitting result in Ref.~\cite{Li} and the intercepts are from -210~s to -350~s. %We see that the deep blue area is included in the intercepts of red area.
In Ref.~\cite{Li} the fitting result comes from the data of 12 events while other 6 events are abandoned, as shown is Fig.~\ref{fig:1b} that 12 points fall on the deep blue area while the rest are not. Combing the results of both photons and neutrinos, we suggest that these neutrinos and photons come from the same time frame of GRBs, and inspired by this we enlarge the intercepts of neutrinos to the red ones, and the result is shown as the light blue area. We see clearly that 5 more points of neutrinos are included in the area, and if the rest one point are considered as valid data, the lower limit of the intercept can be more lower. The consistence of the intercepts of both neutrinos and photons suggests that they come from the same time frame of GRBs, in other words the photon events analysed in this work indirectly support the conclusion of pre-burst neutrino emissions in Ref.~\cite{Li}.

\section{Conclusion}

In this work we search 48 GRBs with redshifts to find pre-burst high energy photon events from FGST data, and analysis on these observed events from FGST suggests that they are related with a 1.07~TeV photon event from MAGIC and a number of near-TeV GRB neutrinos from IceCube. The combination of them reveals a picture that there is a pre-burst stage with a long duration period of several minutes of neutrino emissions accompanied with high-energy photons, within a time interval of about 140 to 820 seconds before the low-energy photon burst of a GRB in the source reference frame.
This picture provides a comprehensive understanding of the observed high energy photon events and neutrino events, and strongly supports the photon and neutrino speed variations suggested from phenomenological analyses.
Of course more detections of high energy photon and neutrino events are still necessary to check the picture revealed in this work.
\\

%\section
\noindent
{\bf{Acknowledgements:}}
 This work is supported by National Natural Science Foundation of China (Grant No.~12075003).





\end{document}